\documentclass[journal]{IEEEtran}

\usepackage{hyperref}
\usepackage{graphicx}
\usepackage{amsmath}
\usepackage{mathtools}
\usepackage{mathrsfs}
\usepackage{multirow}
\usepackage[dvipsnames]{xcolor}
\usepackage{soul}
\usepackage{comment}
\usepackage{caption}
\usepackage{subcaption}
\usepackage{blochsphere}
\usepackage{physics}
\usepackage{tikz}
\usepackage{quantikz}
\usepackage{pgfplots}
\usetikzlibrary{positioning,arrows,calc,math,angles,quotes}

\newcommand{\slac}[1]{{\color{violet}[\textsuperscript{sla}#1]}}
\newcommand{\mtd}[1]{{\color{blue}[#1]}}

\newcommand{\red}[1]{\textcolor{Red}{#1}}

\definecolor{AQAMColor}{rgb}{0.03,0.27,0.49}
\colorlet{QFMColor}{AQAMColor!40}
\colorlet{RAColor}{AQAMColor!20}
\colorlet{QAMColor}{AQAMColor!60}

\ifCLASSINFOpdf
\else

\fi

\begin{document}
%
\title{Depth-Based Matrix Classification for the HHL Quantum Algorithm}


\author{\IEEEauthorblockN{Mark Danza\IEEEauthorrefmark{1},
Sonia Lopez Alarcon\IEEEauthorrefmark{2}, Cory Merkel\IEEEauthorrefmark{3}}\thanks{This material is based upon work supported by the
National Science Foundation under Award}
\\
\IEEEauthorblockA{KGCOE Department of Computer Engineering,
Rochester Institute of Technology\\
Rochester, New York\\
Email: \IEEEauthorrefmark{1}\href{mailto:mtd8140@rit.edu}{mtd8140@rit.edu},
\IEEEauthorrefmark{2}\href{mailto:slaeec@rit.edu}{slaeec@rit.edu},
\IEEEauthorrefmark{3}\href{mailto:cemeec@rit.edu}{cemeec@rit.edu}}}
\maketitle

\author{Mark Danza\KGCOE Department of Computer Engineering\\Rochester Institute of Technology\\Rochester, United States of America\\
\href{mailto:abc8255@rit.edu}{rit.edu} }
\maketitle

\begin{abstract}
Under the nearing error-corrected  era of quantum computing, it is necessary to understand the suitability of certain post-NISQ algorithms for practical problems. One of the most promising, applicable and yet difficult to implement in practical terms is the Harrow, Hassidim and Lloyd (HHL) algorithm for linear systems of equations. 
An enormous number of problems can be expressed as linear systems of equations, from Machine Learning to fluid dynamics. However, in most cases, HHL will not be able to provide a practical, reasonable solution to these problems. This paper's goal inquires about whether problems can be labeled using Machine Learning classifiers as suitable  or unsuitable for HHL implementation when some numerical information about the problem is known beforehand. This work demonstrates that training on significantly representative data distributions is critical to achieve good classifications of the problems based on the numerical properties of the matrix representing the system of equations. Accurate classification is possible through Multi-Layer Perceptrons, although with careful design of the training data distribution and classifier parameters.

\end{abstract}


	\section{Introduction}\label{sec:introduction}
The HHL algorithm by Harrow, Hassidim and Lloyd is a well known quantum algorithm for quantum-mechanically
 constructing the solution of a linear systems of equations \cite{HHL}. HHL is one of those quantum algorithms that will only make sense under quantum error-corrected implementation. Although its depth (number of gate layers) varies depending on certain conditions as it will be shown, HHL results in deep quantum circuits. As we approach this new era of quantum computing, it is necessary to gain understanding of the actual implementability of certain algorithms.
 
 The linear system of equations problem can be defined as, given a matrix $A$ and a vector $\vec{b}$, find a vector $\vec{x}$ such that $A\vec{x} = \vec{b}$. In quantum notation, this is expressed as $A\ket{x}=\ket{b},$  where $A$ is a Hermitian operator --- a workaround exists when $A$ is not Hermitian--- and $\vec{b}$ has to be encoded in a quantum state $\ket{b}$ and, hence, it has to be normalized. The solution to this linear system of equations is, therefore, expressed as $\ket{x}=A^{-1}\ket{b}$. The problem boils down to finding $A^{-1}$, like in any other algorithm to solve linear systems of equations. 
 This, in theory, can be solved rather easily in a quantum-mechanical manner through phase representation and phase estimation (Section \ref{sec:background}. However, this approach is full of pitfalls as it is discussed at length in the paper by Scott Aaronson \cite{aaronson2014quantum}. 

Given the matrix representing the linear system of equations, the magnitude of its eigenvalues and the sparsity of the system are critical for defining the depth and precision of the HHL algorithm implementation.  This algorithm can potentially construct the state of the solution vector in running time complexity \begin{equation}O(\log(N)s^{2}\kappa^{2}/\epsilon) \label{eq:timecom} 
\end{equation}

\noindent given that the matrix $A$ is $s$-sparse and well-conditioned, where $\kappa$ denotes the condition number of
the system, and $\epsilon$ the accuracy of the approximation ~\cite{HHL}. The condition number of $A$ is generally defined as

\begin{equation}
\kappa=\sigma_{max}(A)/\sigma_{min}(A)
\end{equation}

\noindent the fraction of the maximum and minimum singular values of $A$. In the case of a normal matrix, this reduces to

\begin{equation}
\kappa=\abs{\lambda_{max}(A)}/\abs{\lambda_{min}(A)}
\label{eq:kappa}
\end{equation}

\noindent the ratio of the maximum and minimum eigenvalue magnitudes.

 \subsection{The importance of depth.}
\begin{figure}
    \centering
    \includegraphics[width=0.95\linewidth]{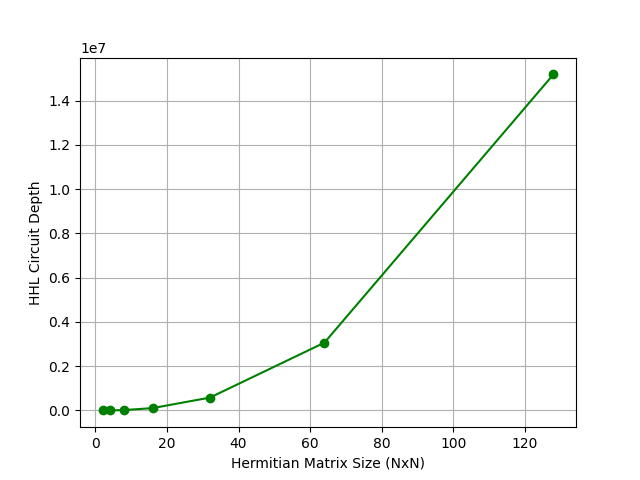}
    \caption{Quantum Circuit Depth vs. Ideal Matrix Size
    }
    \label{fig:ideal-mat-size}
\end{figure}

  The number of layers or depth of a circuit is not only a proxy for execution time, but also a source of accumulated noise that compounds through time and gates, even under error corrected conditions.
Figure \ref{fig:ideal-mat-size} shows the depth growth of the HHL implementation (collected through Qiskit tools \cite{Qiskit}) when A is an  \emph{ideal matrix}. 
We define an ideal matrix as a diagonal matrix in which the diagonal elements (the eigenvalues) are the values 1 and 1/2, repeated along the diagonal. 
These values were chosen for being easily invertible powers of two that can also be represented precisely with few qubits. The condition number for an ideal matrix of any size is $\kappa=2.0$. As it will be shown in Section \ref{sec:background}, in this case the solution to the problem is reduced to its simplest expression ---inverting the eigenvalues--- and the circuit is close to its minimal possible implementation. Despite this, Figure \ref{fig:ideal-mat-size} illustrates that the depth of the circuit grows exponentially, surpassing $1.4\times10^7$ computational steps for the matrix size $128\times128$. The depth growth is much more steep for non-ideal matrix cases. 
Strategies can be explored to improve the implementation of the HHL circuit, such as approximate computing \cite{Crimmins_2024, Camps_2020, sajadimanesh_practical_2022}. Also, it should be noted that the results in Figure \ref{fig:ideal-mat-size}, as it is also expressed in Equation \ref{eq:timecom}, are dependent on the desired accuracy of the solution to the problem expressed as a quantum superposition. This factor could also be adjusted to ease complexity demands on the circuit's implementation. Even with these strategies, for most of the cases the depth of the circuit will be impractical. 

But the number of problems that can be expressed as a linear system of equations is immense. It is possible that among the plethora of problems that can be expressed in this way, some are suitable to be solved through HHL. 

Given the daily advances in quantum computing to handle depth, noise and quantum error correction, it is hard to set a limit to the parameters under which HHL's implementation will be suitable. This paper proposes that \emph{given a certain depth threshold}, matrices representing the linear system of equations can be classified as suitable or non-suitable for HHL through machine learning. As a proof of concept, this work trains and tests on a large number of linear systems of equations, and analyzes the impact of sparsity and condition numbers on the results. As a practical implementation, the approach is tested on the ability to solve the linear system of equations representing a single layer neural network to solve the classification of the IRIS dataset \cite{Iris}.

	\section{HHL-Algorithm Implementation}\label{sec:background}

Many problems can be reduced to solving a linear system of equations, and as such, HHL-based quantum acceleration has been a source of hope in many fields. 
As stated above, in quantum notation this linear system of equations is expressed as 
\begin{equation}A\ket{x}=\ket{b} \label{eq:LinearSystemofEquations} \end{equation}
\noindent where $A$ is a Hermitian operator, and $\vec{b}$ has to be encoded in a quantum state $\ket{b}$ and, hence, it has to be normalized.  Given a matrix $A$ of size $N\times N$, classical algorithms would have a computational complexity $O(N^3)$. If the operator were expressed as a diagonal matrix, the calculation of $A^{-1}$ is almost immediate, creating a diagonal matrix in which the eigenvalues $\lambda_i$ in the diagonal are replaced by their corresponding inverted values, $\lambda_i^{-1}$. The computationally intensive portion of this solution in the classical context would be precisely to calculate the eigenvalues of $A$. This, however, can be solved rather easily in a quantum-mechanical manner through the use of the quantum phase estimation in which the eigenvalues are calculated and expressed as a phase. 

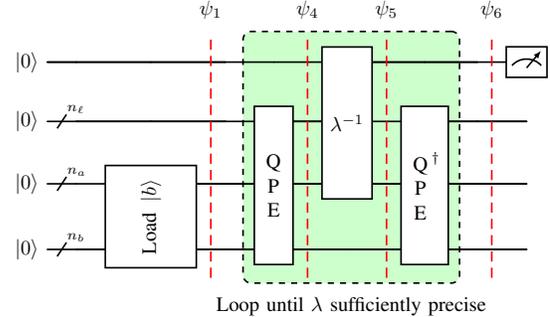
\begin{figure}[h]
    \centering
    \resizebox{0.4\textwidth}{!}{\begin{quantikz}
      \lstick{$\ket{0}$} & \qw & \qw\slice[label style={yshift=0.2cm}]{$\psi_1$} & \qw & \gategroup[4,steps=3,style={dashed,rounded corners,fill=green!20, inner xsep=2pt},background,label style={label position=below,anchor=north,yshift=-0.3cm}]{{Loop until $\lambda$ sufficiently precise}}\slice[label style={yshift=0.2cm}]{$\psi_4$} & \gate[3,disable auto height]{\lambda^{-1}}\slice[label style={yshift=0.2cm}]{$\psi_5$} & \qw & \qw\slice[label style={yshift=0.2cm}]{$\psi_6$} & \meter{} \\
      \lstick{$\ket{0}$} & \qwbundle{n_\ell} & \qw & \qw & \gate[3,disable auto height]{\verticaltext{QPE}} & \qw & \gate[3,disable auto height]{\verticaltext{QPE}^{\dagger}} & \qw & \qw \\
      \lstick{$\ket{0}$} & \qwbundle{n_a} & \gate[label style={black,rotate=90}, wires=2]{\text{Load } \ket{b}} & \qw & \qw & \qw & \qw & \qw & \qw \\
      \lstick{$\ket{0}$} & \qwbundle{n_b} & \qw & \qw & \qw & \qw & \qw & \qw & \qw
    \end{quantikz}}
    \caption{Harrow-Hassidim-Lloyd circuit overview.}
    \label{qiskit_HHL}
\end{figure}

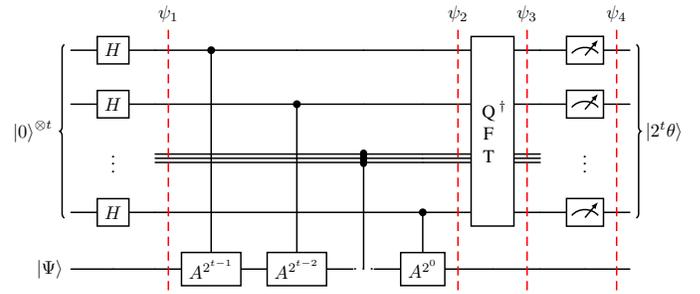
\begin{figure}[h]
    \centering
    \resizebox{0.5\textwidth}{!}{\begin{quantikz}[wire types={q,q,n,q,q}, classical gap = 0.08cm]
      \lstick[4]{$\ket{0}^{\otimes t}$} & \gate{H} & \slice{$\psi_1$} & \ctrl{4} & &&\slice{$\psi_2$}& \gate[4,disable auto height]{\verticaltext{QFT}^{\dagger}}\slice{$\psi_3$} &  & \meter{}\slice{$\psi_4$} & \rstick[4]{$\ket{2^t\theta}$}\\
       & \gate{H} & & & \ctrl{3} & & & & & \meter{} & \\
       & \push{\vdots} &  & \setwiretype{b} &  & \ctrl{0}\wire[d][2]{q} & & &  & \setwiretype{n}\push{\vdots} & \\
       & \gate{H} & & & & & \ctrl{1} & & & \meter{} & \\
       \lstick{$\ket{\Psi}$} & & & \gate{A^{2^{t-1}}} & \gate{A^{2^{t-2}}} & \push{\dots} & \gate{A^{2^0}} & & & &
    \end{quantikz}}
    \caption{Quantum phase estimation (not inverse) circuit overview.}
    \label{tikz:qpe}
\end{figure}

If the matrix $A$ is not Hermitian, the matrix and input vector are modified such that:
\begin{equation}
    \label{dilation} 
    \begin{bmatrix} 0 & A\\ A^\dag & 0 \end{bmatrix} \begin{bmatrix} 0 \\ x \end{bmatrix} = \begin{bmatrix} b \\ 0 \end{bmatrix}
\end{equation}

\noindent which ensures the new matrix is Hermitian. In this case, the size of the matrix is doubled. 
Another approach to setting up the input matrix is to transform the system of linear equations to
\begin{equation}
    A^TA|x\rangle=A^T|b\rangle
\end{equation}
Modifying the input matrix to be $A^TA$ means the input matrix will always be a square matrix, which is a requirement for the HHL algorithm. 

The circuit implementation can be broken down into five different circuit stages: loading the $|b\rangle$
input, Quantum Phase Estimation, Eigenvalue Inversion, inverse Quantum Phase Estimation, followed by measurement as shown in Figure \ref{qiskit_HHL}.  It is important to take into account that the algorithm assumes the case being considered is one in which the solution $\ket{x}$ does not need to be fully known, but rather an approximation of the expectation value of some operator associated with the solution vector.  $\ket{x}$ is in a superposition state of all the independent values of the vector. It is an amplitude encoded solution: the amplitude of each basis state contains each of the weights, the solution to the optimization problem.  Trying to read each of them would require an amount of time such that it would ruin any quantum advantage. 

\begin{figure*}[h]
    \centering
    \begin{subfigure}[b]{0.45\textwidth}
        \centering
        \includegraphics[height=1.5in]{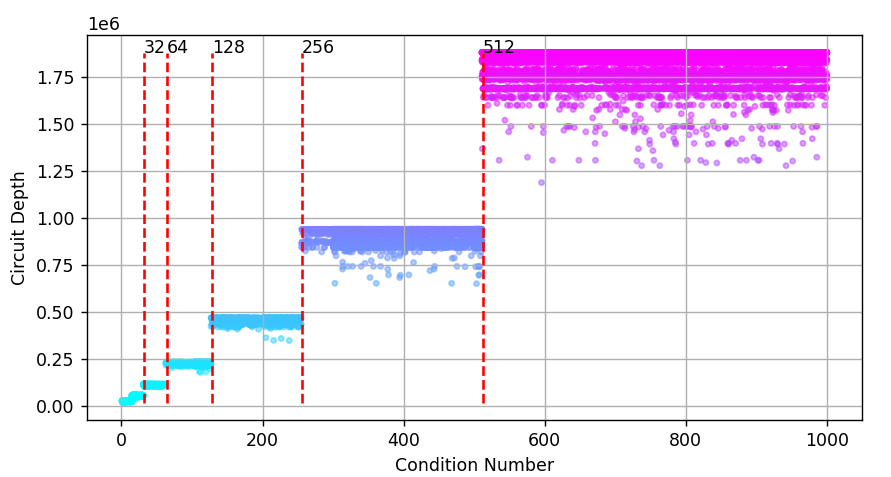}
        \caption{Circuit depth vs. condition number for all sparsities plotted together.}
        \label{fig:skappadepth-2d}
    \end{subfigure}
    ~
    \begin{subfigure}[b]{0.45\textwidth}
        \centering
        \includegraphics[height=2in]{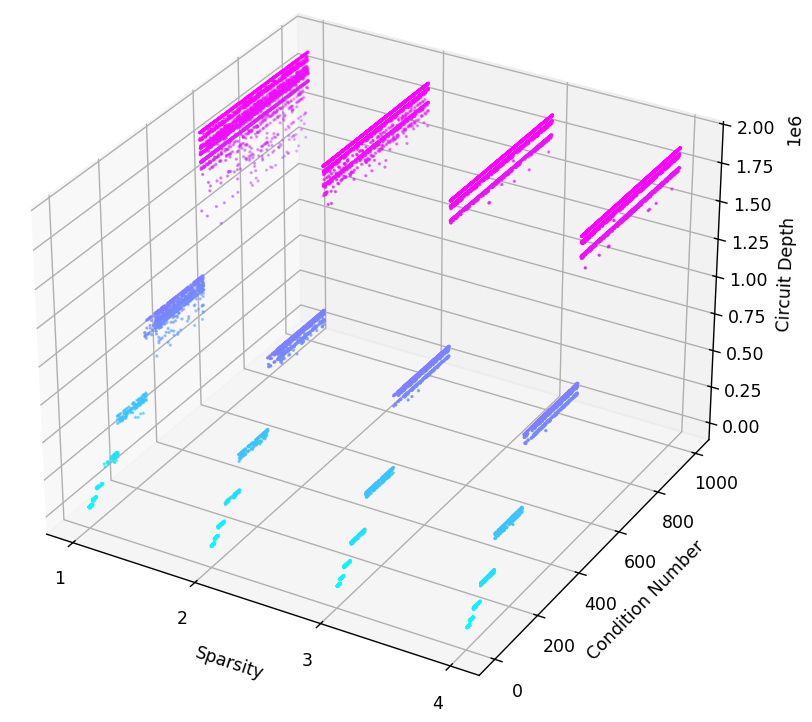}
        \caption{Circuit depth vs. sparsity and condition number.}
        \label{fig:skappadepth-3d}
    \end{subfigure}
    \caption{Distribution of HHL circuit depths for $4\times 4$ matrices. Each point represents a sample corresponding to one matrix in the dataset. Purple points have the highest depth, and light blue have the lowest.}
    \label{fig:skappadepth}
\end{figure*}

\subsection{Circuit's Depth and Time complexity}
 
The HHL algorithm can potentially estimate the function of the solution vector in running time complexity
$O(\log(N)s^{2}\kappa^{2}/\epsilon)$, given that the matrix $A$ is $s$-sparse and well-conditioned, where $\kappa$ denotes the condition number of
the system, and $\epsilon$ the accuracy of the approximation ~\cite{HHL}. In practical terms, the time complexity is reflected in the depth of the quantum circuit. The sparsity of the system's matrix and the condition number are key parameters that are dependent only on the numerical nature of the system of equations. The source of this connection is in the Phase Estimation algorithm. 

The \emph{Phase Estimation (PE)} step, depicted in Figure \ref{tikz:qpe}, extracts the eigenvalues of $A$ onto a PE quantum register. The matrix $A$ is implemented as a controlled operator as many times as there are qubits in the PE register. Therefore, the sparsity and condition number are key sources of computational steps as follows: 
\begin{enumerate}
    \item \emph{sparsity:} the matrix $A$ is encoded as a controlled operator. Encoding these operators requires a number of steps that  grows with the number of elements in the matrix.
    \item \emph{condition number, $\kappa$:} the precision of the eigenvalues in the phase domain is directly related to the size of the PE register, which in turn is related directly to how many times the controlled operator $A$ should be applied. 
\end{enumerate}

Figure \ref{fig:skappadepth} illustrates this dependency. As it will be explained in more detail in Section \ref{sec:proposed}, for this work 58,325 sample matrices ($A$) were generated, and HHL implementations for each one of them with a normalized uniform $\ket{b}$ state were implemented to extract their fully decomposed  quantum circuit depth using Qiskit tools \ref{qiskit_HHL}. Figure \ref{fig:skappadepth-3d} shows the depth depending on  the condition number, $\kappa$ and sparsity, $s$, of $A$. As it can be seen, the depth behaves as a step function on the condition number. This is also depicted in Figure \ref{fig:skappadepth-2d}, where it is clear that steps happen when condition number doubles in size. The reason for this is that the encoding of the eigenvalues is normalized to the smallest of them. The number of qubits necessary to implement the  eigenvalues grows by one when the fraction $\abs{\lambda_{max}(A)}/\abs{\lambda_{min}(A)}$ surpasses each power of two. This increases the depth of the implementation greatly, since a Controlled-$A^{2^{t-1}}$ implementation is included for each new phase estimation most-significant qubit. As sparsity (non-zero elements) grows in Figure \ref{fig:skappadepth-3d}, the depth bands narrow down towards the higher end, since the Hamiltonian simulation of matrices with higher numbers of non-zero elements are more likely to require more steps. 

\subsection{Related Work}\label{sec:HHL}

L\'opez Alarc\'on et al. proposed the use of the swap test to calculate the distance between a known test point and the actual solution to the training problem  \cite{ICCD2022}. Other hybrid approaches have been proposed  \cite{lee_hybrid_2019, yalovetzky_hybrid_2023} 
 with the goal of reducing the depth of the design by including an intermediate classical step that takes in  a Quantum Phase  Estimation (QPE) of the matrix's eigenvalues and generates a reduced ancillary quantum encoding circuit (AQE) for the continuation of the algorithm. 
Based on the HHL solution to a linear regression problem \cite{Dutta2020},  Zhao et al. proposed a solution to Bayesian deep learning using a quantum computer \cite{bayesiandl}. This example was a proof of concept on a small, pre-set system of equations. 
The present work, however, does not look at optimizing the circuit implementation  to reduce depth, but it focuses on the impact of the numerical nature of the problem at hand on the depth of the circuit description. 
  Scott Aaronson in his \emph{``Quantum Machine Learning: Read the fine print''} \cite{aaronson_read_2015}  discusses in detail all the limitations of the HHL algorithm, from the precision of the solution to the sparsity of the input matrix. Previous work has made efforts to use machine learning regression to predict matrix condition numbers using precomputed features extracted from the matrix data \cite{xu_new_2004}, \cite{han_comparison_2007}. These will all be carefully considered and evaluated in this work.

\section{Matrix Classification for HHL}\label{sec:proposed}

Equation \ref{eq:timecom} shows that the time complexity (or in practical terms, the depth) depends quadratically on the size, sparsity and the condition number of the matrix $A$. In theory, given these three, a decision could be made of whether this matrix is a good fit for HHL or not. The sparsity, $s$ can be estimated somewhat easily when $A$ is known. But the condition number, $\kappa$, (see Equation \ref{eq:kappa}) depends on the eigenvalues, and classically calculating the eigenvalues is as complex as solving the system of equations. Figure \ref{fig:skappadepth} shows that $\kappa$ is in fact a stronger predictor of depth than $s$, but it is the hardest of the two to calculate.

In this work, we  use Machine Learning to learn about the depth of HHL implementations depending on the system's matrix $A$. Machine learning offers promising techniques for obtaining a function (i.e., a model) that maps samples from a high-dimension input space to a limited set of well-understood outputs. The problem of determining whether a given matrix can be efficiently implemented in a quantum HHL circuit may be framed as a binary classification problem which can be solved in this way, to a certain degree of accuracy. In this case, the problem is set as a binary classification problem in which matrices will be labeled as \emph{well suited} or \emph{poorly suited} for HHL. Supervised learning requires a large amount of data to be collected and labeled for training of the classification engine -- in this case, a variety of valid input matrices must be gathered and identified as efficiently implementable in a quantum HHL solution or not.

Determining the criteria that makes a matrix suitable or not is challenging. In the current state of quantum hardware, no matrix $A$ is well suited for HHL. Simply by looking back at Figure \ref{fig:ideal-mat-size} it can be seen that thousands of layers are needed for small matrices that are already diagonalized and expressed as powers of 2. Thousands of layers with current quantum computing technologies is a non-implementable problem, but as technology advances, deeper, error-corrected, and more resilient implementations will be possible.  The work presented in this document does not state what the depth criteria must be to make this classification, but intends to find out if \emph{given a classification criteria based on depth,} the ML model can learn if the matrix falls on one side or the other of this criteria.

In addition, it should be noted that the $\vec{b}$ vector is an important part of any linear system of equations, and the state preparation of $\ket{b}$ adds additional layers to the HHL implementation. The preparation of $\ket{b}$ depends entirely on the nature of $\vec{b}$. All linear systems of equations considered in this work use an efficiently implementable $\ket{b}$ that has a consistent and negligible impact on the depth of the HHL circuit. The complexity of implementing $\ket{b}$ is a challenge of state preparation and is outside the scope of this work.

\subsection{Producing Datasets for Training and Testing}\label{data-collection}

The matrix input to an HHL problem is an $N\times N$ Hermitian matrix, where $N$ is a power of 2. To accommodate a greater diversity of problems, it is assumed that any non-Hermitian matrix is transformed into a Hermitian matrix by dilation, as in Equation \ref{dilation}. 
To generate the training sets, 58,325 matrices in sizes ranging from $2\times 2$ to $16\times 16$ are generated randomly with a certain sparsity condition. Sparsity of these matrices range from 1 to the maximum possible, equal to the dimension $N$ for each matrix size. Each random matrix is normalized by dividing each matrix element by the matrix norm, such that the singular values have a magnitude between $1/\kappa$ and $1$, as assumed in \cite{HHL}. The condition number $\kappa$ is calculated for each random matrix, and only  
matrices with a condition number below 1000 are kept, under the presumption that matrices that are much more ill-conditioned than that are very unlikely to have an efficient quantum implementation, as it will be shown. To isolate the impact of the matrix on the quantum implementation, every system of equations is solved for an  appropriately sized $\ket{b}$ (Equation \ref{eq:LinearSystemofEquations}) in uniform superposition (normalized vector of ones), and potentially dilated as part of the HHL algorithm. Because of this, the $\ket{b}$ state preparation portion of the circuit will contribute the same amount of depth to every circuit of a given problem size. 

To motivate the assignment of training labels, a Qiskit utility for linear system solvers that supports HHL \cite{vazquez_anedumlaquantum_linear_solvers_2025} is used to algorithmically construct HHL circuits for the randomly generated input matrices. The depth of the entire HHL implementation (in elementary gates) is measured and used to quantify the implementation efficiency of different input matrices relative to one another. Since Qiskit's notion of depth relies on the level of decomposition of a given quantum circuit, the HHL circuit is repeatedly decomposed until its measured depth becomes constant to attain the "full" circuit depth. By default, the generated HHL circuits use Qiskit's exact reciprocal implementation of conditioned rotations for the eigenvalue inversion portion of the circuit (see Figure \ref{qiskit_HHL}) \cite{vazquez_anedumlaquantum_linear_solvers_2025, noauthor_textbooknotebooksch-applicationshhl_tutorialipynb_nodate}.

\begin{figure}
    \centering
    \includegraphics[width=0.95\linewidth]{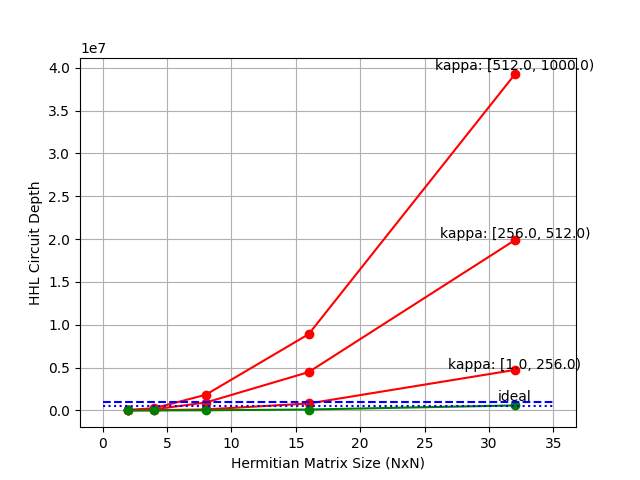}
    \caption{Average HHL circuit depth vs. random matrix size for three $\kappa$ bins and the ideal matrices. The number of layers in the exact quantum circuit grows exponentially with the input size, at faster rates for higher condition numbers. The horizontal blue lines depict the potential depth cutoffs for suitable/non-suitable problems.}
    \label{fig:random-mat-size}
\end{figure}

Figure \ref{fig:random-mat-size} shows how the average HHL circuit depth grows with the size of the random input matrices. The ideal matrix data is the same as the line plotted in Figure \ref{fig:ideal-mat-size}. The other three solid curves in red correspond to subsets of the randomly generated matrix dataset. 
The circuit depths of the samples in a $\kappa$ bin are averaged at each value of $N$. 
The blue lines in Fig. \ref{fig:random-mat-size} represent the depth cutoff boundaries used for matrix classification, described later in this section.

As it is shown in Figures \ref{fig:skappadepth} and \ref{fig:random-mat-size}, the depth of the implementation grows with the condition number and sparsity. However, the condition number has a greater influence in the growth of the circuit. Unfortunately, the condition number is the hardest of the two to calculate or estimate.

This work does not intend to state what the right boundary is for the classification of suitable/not-suitable HHL problems, but to learn about the numerical nature of the problem and the potential of ML classification of these problems for future systems. It is assumed that future quantum systems might have a constant depth limit to be able to generate and output with  reasonable state coherence and tolerance to accumulated quantum noise. For this work, it is presumed that such a boundary exists in the region of the depth data shown in Figure \ref{fig:random-mat-size}. The initial depth cutoff is the constant function
 $   f(A) = 1,000,000$ layers.
The dashed blue line in Figure \ref{fig:random-mat-size} shows where this boundary is located. Samples with a circuit depth below this value are considered well suited, while samples with greater depth are labeled as poorly suited for HHL. This boundary splits the data in Figure \ref{fig:random-mat-size} into 47.6\% well suited (positive samples) and 52.4\% poorly suited (negative samples), which is well balanced for ML training. An alternative depth cutoff is also defined for the purpose of comparison:
    $f(A) = 500,000$ layers.
The alternative boundary is the dotted blue line in Figure \ref{fig:random-mat-size}, just slightly below the original depth cutoff. 
The alternative cutoff results in 36.2\% of the data being positive samples and 63.8\% negative samples. An unbalanced dataset makes it more challenging to assess the accuracy of the ML model. Aiming to train and test on a balanced dataset, this work used the first cutoff function $   f(A) = 1,000,000$, since this will facilitate interpreting the accuracy of the approach. An unbalanced dataset will be analyzed in more detail in Section \ref{sec:IRIS}.

\subsection{Matrix Feature Selection}\label{features} 

Given the set of random matrices generated for training, there are several options on what type of features can be extracted from these in order to train the supervised learning models mentioned above. These are the features that would then have to be extracted in test cases for classification. The extraction of features has a computational cost. Ideally, the learning models would be able to learn from the raw data (input matrix) without the need to extract any features. However, for this research, it is interesting to investigate which features are key in driving good classification results and which features can be more easily learned from the raw data by the learning models. For that reason, four different datasets are explored in this work, each of them based on different manipulations of the raw data in addition to the raw data itself.
Some of the matrix features are taken from \cite{xu_new_2004}, and others are introduced here. In total, 104 features are extracted from the matrix data. These features are organized into several categories based on the nature of the information they capture. The categories are summarized below, and the complete definitions of all features can be found in the Appendix.

\begin{table}
    \centering
    \begin{tabular}{cc}
        \textbf{dataset 1} & extracted features + condition number\\
         \hline
        \textbf{dataset 2} & extracted features + estimated condition number\\
         \hline
        \textbf{dataset 3} & extracted features only\\
        \hline
         \textbf{dataset 4} & raw data ($4\times 4$ matrices)\\
        \hline
       
    \end{tabular}
    \caption{Features used by each dataset for ML training. The datasets use identical matrices, but differ in the features from those matrices that are used to train the classifiers. \emph{dataset 1} provides the most specific information about the condition number. Less conditioning information is provided in \emph{dataset 2}, even less in \emph{dataset 3}, and the least in \emph{dataset 4}.}
    \label{tab:datasets}
\end{table}

\subsubsection{Structure Based Features} These features are based on the distribution of nonzero elements in the matrix. They include different measures of matrix sparsity, the number of non-zero elements (NNZ), and the symmetry of the matrix (with respect to the main diagonal).

\subsubsection{Value Based Features} These features relate to the distribution of element values in the matrix. They incorporate many statistical measures of the value distribution in different parts of the matrix, as well as several matrix norms.

\subsubsection{Diagonal Based Features} Features of the diagonal represent additional information about the diagonal elements, the distance relationships of other matrix elements with the main diagonal, and the bandwidth of the matrix. Here, "bandwidth" refers to the clustering of non-zero elements about the main diagonal of the matrix.

\subsubsection{Condition Number Features} These features include estimations of the matrix eigenvalues and condition number using techniques related to Gershgorin disks \cite{garren_bounds_1968}. The actual condition number may also be used as a feature for the purpose of comparing model performance with different feature sets, but not in any practical application.



A particular model may be trained using all of these features or a subset of them for comparative analysis. 
Through the training experiments, it was found that as expected the information related to the condition number is key at making predictions on the depth of the circuit. Therefore, the results are shown and discussed based on the presence or absence of condition number information, whether exact calculation, estimated or none.  Table \ref{tab:datasets} summarizes the features included in each dataset. \emph{dataset 1} uses features extracted from a matrix sample, 
including the \emph{condition number}. \emph{dataset 2} is similar to \emph{dataset 1}, but 
includes the \emph{Gershgorin disks estimation} and \emph{Cassini ovals estimation}  of the condition number rather than the exactly calculated condition number. \emph{dataset 3} uses extracted features but does not include any specific condition number information. Last, \emph{dataset 4} consists of only raw matrix data, limited to the $4\times4$ cases, due to computational cost. A model trained on this dataset is only responsible for classifying $4\times 4$ matrices. 



\subsection{Classifier Implementation}\label{modelimpl}




Several binary classification models were explored, and the mult-ilayer perceptron (MLP) was found to provide most favorable combination of accuracy and flexibility. The classifier is an MLP consisting of five fully connected hidden layers with rectified linear unit (ReLU) activation. The first layer has 512 neurons, and the others each have 256. An initial learning rate of 0.01 and momentum of 0.9 are used with stochastic gradient descent optimization. The learning rate starts at 0.01 and adapts to the training process -- if model performance does not increase significantly for 15 consecutive training iterations, the learning rate is divided by 5. Otherwise, the training may be allowed to stop early to alleviate overfitting. These hyperparameter values were obtained first via grid search using scikit-learn tools \cite{noauthor_scikit-learn_nodate}, then tuned further by hand.


\subsection{Evaluation Metrics}\label{modeleval}

Confusion matrices and the various performance metrics derived from them are used to evaluate and compare models with regard to predictions on the test data. In particular, a model's \emph{accuracy} is defined as the fraction of all correct predictions. Given that the training and testing is approximately balanced (as many positive as negative samples), this is a good first metric of the behavior of the models. \emph{Recall} is the proportion of positive samples that are predicted as positive. Similarly, \emph{specificity} is the proportion of negative samples predicted correctly. The precision of a model is the fraction of positive predictions that are correct. 
The \emph{F1 score} is the harmonic mean of recall and precision, calculated as:

\begin{equation}F1 = \frac{2\times precision\times recall}{precision + recall}\end{equation}

\emph{F1 score} is generally a more holistic measure of model performance than accuracy. It captures two metrics (recall and precision) that may be at odds if the model does not generalize well. If either precision or recall are low, the F1 score will suffer. For all the performance metrics described in this section, the possible values range from 0 to 1 (or they may be represented as a percentage). A reliable model scores as close to 1 as possible in every metric.

        \section{Results with generic training}\label{sec:results}


A separate MLP classifier was trained on each dataset presented in Table \ref{tab:datasets}; the performance metrics on the test data are collected in Table \ref{tab:scores}. As a reminder, \emph{datasets 1-3} progressively let go of condition number information, while \emph{dataset 4} has its condition number information buried in the actual matrix, since this dataset is referring to raw data. As expected, \emph{datasets 1-3} make progressively worse predictions, while \emph{dataset 4} has the most difficulty at making correct classifications based on purely raw data. When the exact calculation of the condition number is provided, the model is overwhelmingly accurate in all metrics. As shown in Figure \ref{fig:skappadepth}, the clear step function on $s$ and $kappa$ makes depth reasonably predictable when these parameters are known.  Calculating the condition number is computationally as complex as solving the linear system of equations, and it is not proposed as a viable option, but it sets a good ground truth and insight to understand the rest of these results.  As the exact information on the condition number is progressively lost (\emph{datasets 2-3}), the metrics get slightly worse. In particular, the models struggle most with predicting a problem as non-suitable for HHL when it is in fact suitable, as indicated by the lower recall metric. Last, training on the raw data would be the ideal case scenario, since it avoids the computational complexity of extracting features, or calculating or estimating the condition number. However, this is the case with lowest accuracy and overall metrics. Only 41\% of suitable problems will be classified as such, while it will accurately predict when a problem is too deep in 84\% of cases. Therefore, this classification model will err on the side of caution when asked if a problem is suitable or not for HHL.   

\begin{figure}
    \centering
    \includegraphics[width=0.95\linewidth]{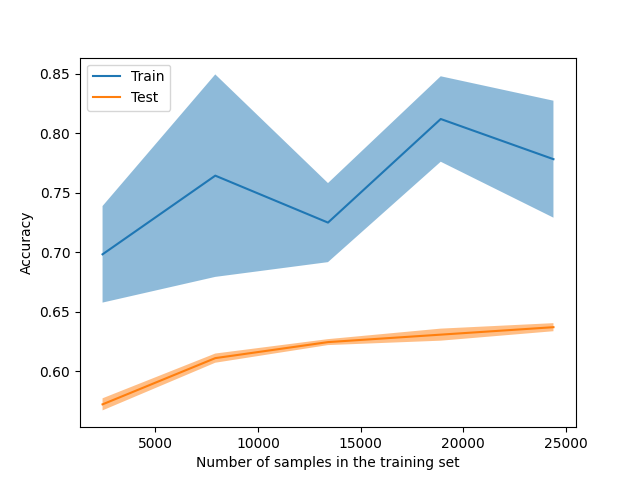}
    \caption{Learning curve with five-fold cross validation for the \emph{dataset 4} (raw data) classifier. 
    The blue line (top) shows how the model's training accuracy varies with the number of samples allotted for training. The orange line (bottom) shows the same for test accuracy.}
    \label{fig:learning-curve}
\end{figure}

The performance of the model for \emph{dataset 4}  may be dependent on the number of training samples. Generally speaking, training on raw data requires a much larger number of samples, since the expectation is that the ML model will ``learn'' the features rather than having them extracted for it.  
To determine whether this is the case, Figure \ref{fig:learning-curve} shows a learning curve plot relating the model accuracy to the amount of data used for training. The learning curve is constructed using five-fold cross validation. \emph{Cross validation} is performed by splitting the data into $k$ partitions, or \emph{folds}, setting aside one partition for validation, and using the remaining partitions for training. This process is repeated, using each fold in turn as the validation set, and the results from all $k$ models are averaged together. In Figure \ref{fig:learning-curve}, the solid lines represent the average training and test accuracy, and the shaded regions capture the variance across the different folds. The training accuracy (blue, top) shows a positive trend but is highly variable. The test accuracy (orange, bottom) has a much tighter positive trend; however, its rate of increase slows as the number of samples grows. The test accuracy reaches a plateau in this graph, indicating that using more training data will not appreciably increase the model performance. This demonstrates the limitations of the raw data model.

\begin{table}
    \centering
    \begin{tabular}{|l|cccc|}
    \hline
    \multicolumn{1}{|c|}{\multirow{2}{*}{\textbf{Metric}}} & \multicolumn{4}{c|}{\textbf{Dataset}}                                                                                   \\ \cline{2-5} 
    \multicolumn{1}{|c|}{}                                 & \multicolumn{1}{l}{1}              & \multicolumn{1}{l}{2}     & \multicolumn{1}{l}{3}     & \multicolumn{1}{l|}{4}     \\ \hline
    accuracy                                               & \textit{0.992}                     & 0.781                     & 0.749                     & 0.643                      \\
    F1 score                                               & \textit{0.992}                     & 0.753                     & 0.693                     & 0.517                      \\
    recall                                                 & \multicolumn{1}{l}{\textit{0.991}} & \multicolumn{1}{l}{0.691} & \multicolumn{1}{l}{0.586} & \multicolumn{1}{l|}{0.415} \\
    specificity                                            & \textit{0.993}                     & 0.864                     & 0.901                     & 0.838                      \\ \hline
    \end{tabular}
    \caption{Classifier scores for each dataset. Results of the model that trained with the actual condition number are italicized -- note that those are for comparison only and computing the actual condition number as a data feature is too impractical for real-world applications.}
    \label{tab:scores}
\end{table}




\subsection{Alternate Depth Cutoff}

The same data were relabeled according to $f(A) = 500,000$ layers to represent a different cutoff for the acceptable HHL circuit depth, resulting in a new distribution of training data. The scores of the retrained models are listed in Table \ref{tab:scores-new}. Consider these results as compared to Table \ref{tab:scores}. For \emph{datasets 2-4}, there is a higher accuracy than in Table \ref{tab:scores}, and this is supported by higher F1 scores in the case of \emph{datasets 2 and 3}. In contrast, dataset 4 exhibits a lower F1 score, showing that the raw data model struggles to learn what characterizes a matrix as "well suited" when there are fewer positive examples. The higher specificity scores across \emph{datasets 2-4} show that the new models are more adept at classifying negative samples, which is to be expected given that there are more negative samples to learn from. The performance boost of the \emph{dataset 3} model is particularly interesting -- very similar in all metrics to the model for \emph{dataset 2}. Previously, the \emph{dataset 3} model was impacted substantially by the lack of specific condition number information. This information becomes less important when the condition on depth is more strict, and the decision can be more easily be made based on other matrix features. In particular, looking back at Figure \ref{fig:skappadepth}, sparsity may be informative to make depth classifications when the limit is set low enough.  

\begin{table}
    \centering
    \begin{tabular}{|l|cccc|}
    \hline
    \multicolumn{1}{|c|}{\multirow{2}{*}{\textbf{Metric}}} & \multicolumn{4}{c|}{\textbf{Dataset}}                                                                                   \\ \cline{2-5} 
    \multicolumn{1}{|c|}{}                                 & \multicolumn{1}{l}{1}              & \multicolumn{1}{l}{2}     & \multicolumn{1}{l}{3}     & \multicolumn{1}{l|}{4}     \\ \hline
    accuracy                                               & \textit{0.994}                     & 0.848                     & 0.842                     & 0.755                      \\
    F1 score                                               & \textit{0.992}                     & 0.768                     & 0.754                     & 0.467                      \\
    recall                                                 & \multicolumn{1}{l}{\textit{0.988}} & \multicolumn{1}{l}{0.681} & \multicolumn{1}{l}{0.653} & \multicolumn{1}{l|}{0.329} \\
    specificity                                            & \textit{0.998}                     & 0.947                     & 0.953                     & 0.961                      \\ \hline
    \end{tabular}
    \caption{Classifier scores for each dataset with the alternative training labels.}
    \label{tab:scores-new}
\end{table}


\section{Iris Dataset Predictions}
\label{sec:IRIS}
The main goal of the trained models is to determine whether a particular problem space of linear systems of equations holds potential for quantum acceleration using HHL. Take the iris flower problem for example. The iris dataset comprises 150 samples with four features and is commonly used for ML classification \cite{Iris}. By selecting four linearly independent samples, a linear system of equations is formed where the samples are the rows of the coefficient matrix. Such a matrix may be used to train a simple model for least squares classification, which involves a matrix inverse. If this is done as a subroutine in a larger quantum computation, HHL may be used to perform the matrix inversion. To this end, one iris matrix could be evaluated using the classifiers from Table \ref{tab:scores} to predict whether HHL would provide quantum advantage over a classical solution. However, to get a broader idea of whether HHL should be applied to the iris problem space, many iris matrices can be evaluated to estimate the proportion that are well suited. In this case, 8301 invertible matrices are randomly drawn from the iris dataset. The associated circuit depths of these matrices are found and used to assign ground truth labels by the same process described in Section \ref{data-collection}. Approximately 80\% of the iris matrix samples are labeled as well suited for HHL according to our depth based criteria. 

\subsection{Generically Trained Classifiers on the Iris Dataset}

Treating the iris matrices as a set of test samples, the classifiers were reevaluated. 
Table \ref{tab:scores-iris} shows the scores related to the iris test matrices. 
The results here show that the \emph{dataset 1} model continues to outperform the other classifiers, which struggle to make accurate predictions on the iris data. In particular, the \emph{datasets 3-4} classifiers have very low accuracy, F1, and recall scores. This reveals that these models have difficulty classifying all the positive samples, causing each of these metrics to drop since the majority of iris matrices are well suited. 
The key difference between the random training matrices and the iris matrices is that the randomly generated matrices lack structure, whereas the iris data is composed of values with specific relationships. The iris dataset is not a proper subgroup of the randomly generated samples that were generated for the classifiers described in Section \ref{sec:results}. Two differences are notorious: (i) \emph{sparsity} ($s$) is always 4 --- the highest it can be. This means that the Iris datasets do not have any zero entries; and (ii) \emph{condition numbers} of this set are distributed differently than the randomly generated numbers, even when only the sparsity $s=4$ cases are considered. This is reflected in Figure \ref{fig:hist}. In summary, the training and testing are taking place on very different data distributions. When condition number information is lost, the trained models are no longer useful. 

\begin{table}
    \centering
    \begin{tabular}{|l|cccc|}
    \hline
    \multicolumn{1}{|c|}{\multirow{2}{*}{\textbf{Metric}}} & \multicolumn{4}{c|}{\textbf{Dataset}}                                                                                   \\ \cline{2-5} 
    \multicolumn{1}{|c|}{}                                 & \multicolumn{1}{l}{1}              & \multicolumn{1}{l}{2}     & \multicolumn{1}{l}{3}     & \multicolumn{1}{l|}{4}     \\ \hline
    accuracy                                               & \textit{0.978}                     & 0.557                     & 0.308                     & 0.339                      \\
    F1 score                                               & \textit{0.986}                     & 0.672                     & 0.281                     & 0.336                      \\
    recall                                                 & \textit{0.979}                     & 0.571                     & 0.170                     & 0.210                      \\
    specificity                                            & \multicolumn{1}{l}{\textit{0.971}} & \multicolumn{1}{l}{0.499} & \multicolumn{1}{l}{0.847} & \multicolumn{1}{l|}{0.842} \\ \hline
    \end{tabular}
    \caption{Classifier scores on the iris matrices. These scores pertain to the classifiers trained on all randomly generated matrices with the original depth boundary.}
    \label{tab:scores-iris}
\end{table}

\subsection{Modified Classifiers on the Iris Dataset}
\begin{figure}
    \centering
    \includegraphics[width=0.95\linewidth]{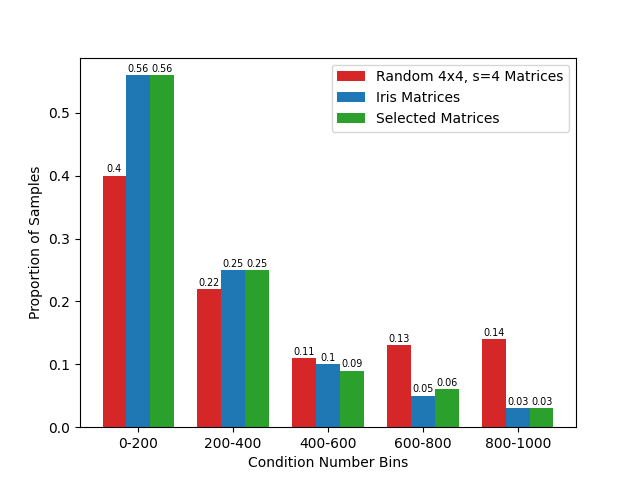}
    \caption{Distribution of condition numbers in the set of random $4\times 4$ matrices with $s = 4$ (left, red), the set of iris matrices (middle, blue), and the set of random matrices selected to match the iris distribution (right, green). For each set of matrices, the proportions sum to 1.0. Each tick on the x-axis represents one of five condition number "bins" up to 1000.}
    \label{fig:hist}
\end{figure}

To further explore this issue, the MLP classifiers are retrained using a set of matrices specifically chosen to mimic the distribution of the iris matrices in terms of sparsity and condition number. To form this training set, the randomly generated matrices with size $4\times 4$ and $s = 4$ are isolated. Then, the distribution of $\kappa$ in the iris matrices is measured by counting the number of iris matrices in each of five "condition number bins" that evenly divide the interval from 1 to 1000. Finally, matrices are drawn from the original random training set in groups proportional to the number of samples in each $\kappa$ bin. The resulting set of matrices, called the \emph{selected matrix set} or \emph{iris-like matrix set}, has a $\kappa$ distribution matching the iris matrices. This is shown by the histogram in Fig. \ref{fig:hist}, where the $\kappa$ distribution in the set of selected matrices is seen to follow the iris matrices very closely.

The classifiers were retrained after extracting training samples from the set of matrices selected to form an iris-like distribution. For all the MLP models, the output layer is a single neuron that outputs a confidence score between 0 and 1. This confidence score is used to make the prediction; by convention, if it is above 0.5 the predicted label is positive (well suited), and the prediction is negative otherwise. However, that threshold value may be adjusted if it is beneficial to the classification task. In the case of the new dense $4\times 4$ matrix classifiers, the threshold value that yields the highest balanced accuracy score is determined algorithmically during training. The \emph{balanced accuracy} is the fraction of positive samples predicted correctly averaged with the fraction of negative samples predicted correctly (also known as the average recall of all classes). 
This technique is necessary to prevent the models from learning to make only positive predictions, an issue caused by the imbalance in the iris-like matrix set (which is about 80\% well suited like the real iris data).

Table \ref{tab:iris-like} summarizes the performance of these models on both the source data (the random matrices with an iris-like distribution) and the target data (the actual iris matrices). The rows in the "validation" section show the classifier scores on a subset of the source distribution that was set aside for testing and excluded from training. Similarly, the "test" section indicates the model scores on the iris matrices. The new classifiers achieve a comparable accuracy and better F1 score than the original classifiers on their respective source distributions. This is not a direct comparison because the two classifier versions are trained using samples from different distributions, but it shows that the same model structure and training process can be applied with similar success using slightly different matrix data. Excluding \emph{dataset 1}, the test predictions show more promise with the new classifiers than the scores in Table \ref{tab:scores-iris}. For \emph{dataset 2}, the recall and specificity were very close in Table \ref{tab:scores-iris}, but diverge significantly in the test section of Table \ref{tab:iris-like}. Meanwhile, the \emph{dataset 3-4} models had much higher specificity than recall when trained on the generic matrices, but this trend is reversed for the test in Table \ref{tab:iris-like}. This shows that the models' abilities to identify well suited samples is improved significantly by training on the iris-like matrix distribution. Furthermore, by selecting a prediction threshold that optimizes balanced accuracy, the accuracy and F1 scores of the iris predictions are markedly higher than for the random data classifiers, though at the expense of specificity. The models' better positive predictions come at the cost of their ability to predict negative samples. This makes sense because we intentionally trained the models on a sample set chosen to mimic a target distribution lacking in negative examples. The \emph{dataset 4} classifier presents the highest accuracy and F1 seen in Table \ref{tab:iris-like}, seemingly due to its propensity for positive predictions. However, the specificity is greater than zero, which means that the classifier did learn some way to determine poorly suited samples, which was the goal of optimizing on the balanced accuracy metric. Moreover, the behavior of the raw data classifier when trained on the iris-like matrices is preferable to the performance observed when trained on the generic matrices, where it shows a strong negative bias (see Table \ref{tab:scores-iris}). Overall, these results demonstrate that the classifier's ability to learn from the raw matrix data improves dramatically when the distribution of training data more closely matches the target data.

\begin{table}[]
    \centering
    \begin{tabular}{|cl|cccc|}
    \hline
    \multicolumn{2}{|c|}{\multirow{2}{*}{\textbf{Metric}}}           & \multicolumn{4}{c|}{\textbf{Dataset}}                                                                                   \\ \cline{3-6} 
    \multicolumn{2}{|c|}{}                                           & \multicolumn{1}{l}{1}              & \multicolumn{1}{l}{2}     & \multicolumn{1}{l}{3}     & \multicolumn{1}{l|}{4}     \\ \hline
    \multicolumn{1}{|c|}{\multirow{4}{*}{Validation}}  & accuracy    & \textit{0.991}                     & 0.761                     & 0.769                     & 0.854                      \\
    \multicolumn{1}{|c|}{}                             & F1 score    & \textit{0.995}                     & 0.858                     & 0.862                     & 0.920                      \\
    \multicolumn{1}{|c|}{}                             & recall      & \textit{0.991}                     & 0.816                     & 0.823                     & 0.960                      \\
    \multicolumn{1}{|c|}{}                             & specificity & \multicolumn{1}{l}{\textit{0.993}} & \multicolumn{1}{l}{0.362} & \multicolumn{1}{l}{0.368} & \multicolumn{1}{l|}{0.072} \\ \hline
    \multicolumn{1}{|c|}{\multirow{4}{*}{Test (iris)}} & accuracy    & \textit{0.874}                     & 0.652                     & 0.593                     & 0.773                      \\
    \multicolumn{1}{|c|}{}                             & F1 score    & \textit{0.926}                     & 0.777                     & 0.721                     & 0.871                      \\
    \multicolumn{1}{|c|}{}                             & recall      & \textit{1.000}                     & 0.764                     & 0.662                     & 0.961                      \\
    \multicolumn{1}{|c|}{}                             & specificity & \multicolumn{1}{l}{\textit{0.381}} & \multicolumn{1}{l}{0.216} & \multicolumn{1}{l}{0.327} & \multicolumn{1}{l|}{0.043} \\ \hline
    \end{tabular}
    \caption{$4\times 4$ dense matrix classifier scores on both the source (validation) and target (test) distributions. These classifiers were trained using only matrices with $N = 4$, $s = 4$, and a $\kappa$ distribution matching that of the iris matrices. Validation scores pertain to data from the source distribution, while test scores are from testing on iris matrices.}
    \label{tab:iris-like}
\end{table}

        \section{Conclusion}\label{sec:conclusion}

The goal of this paper is to learn if, \emph{given a certain circuit depth criteria,} a Machine Learning classification model can be used to discern whether a linear system of equations (expressed as $A\times\ket{x}=\ket{b}$) is suitable or not suitable for HHL implementation. The work has focused on the nature of the system's matrix $A$ and has set aside other parameters that also have an impact on the implementability of HHL circuits such as the state preparation of $\ket{b}$ or the read out of the state $\ket{x}$. 

Based on the numerical nature of $A$, its condition number has a critical impact on the depth of an HHL implementation of this matrix.  This is reflected in the ML models, given the overwhelming success of the models trained with this specific number at making classification decisions. But extracting the condition number out of a matrix is just as computationally complex as solving the linear system of equations. Therefore, the goal of these ML models is to learn from a large set of matrices to extract information and potentially classify new matrices as suitable/not-suitable for HHL given a depth condition. Through  proof of concept on random training cases it is concluded that the absence of condition number information lowers the accuracy of the classification of new general problems. Still, the accuracy in the raw data training is of about 64\% correct classifications, and the ML model will correctly classify 84\% of the non-suitable circuits, showing that it will tend towards conservative classifications against the use of HHL.

For verification on a real problem, the Iris dataset was used. In this case, it is clear that having a carefully tailored training set that resembles this general family of matrices is important to generate accurate classifications. When the system was trained on randomly generated matrices with a similar distribution of condition numbers, the test Iris samples were accurately classified in 77\% of the cases, and with F1 score of 0.87.

In conclusion, ML classification can extract  information out of the datasets when trained on the right distribution of sample matrices. Without knowledge of the condition number, the ML model still can make accurate predictions that would at times err towards avoiding HHL.






%

\bibliographystyle{IEEEtran}

\appendix
\section{Machine Learning Features}\label{FeatureAppendix}

\subsection{Structure Based Features}
\begin{itemize}
    \item \emph{sparsity:} $s$, where a matrix is said to be $s$-sparse if each row contains at most $s$ non-zero elements.
    \item \emph{sparsity-to-size ratio:} The ratio $s/N$, where $s$ is defined as above and $N$ is the matrix size (i.e., an $N\times N$ matrix).
    \item \emph{number of non-zero elements:} The total number of non-zero elements in the matrix.
    \item \emph{non-zero rate (fill rate):} The fraction of non-zero elements in the matrix.
    \item \emph{statistical features of row and column structure:} Minimum, maximum, mean, and standard deviation of the number of non-zero elements per matrix row, and the same for matrix columns.
    \item \emph{non-void diagonals:} The number of matrix diagonals with at least one non-zero element.
    \item \emph{symmetry:} Whether the matrix is symmetric (i.e., $A = A^T$ for a matrix $A$).
    \item \emph{relative symmetric rate:} A ratio equivalent to the number of "matching" non-zero elements divided by the total number of non-zero elements. An element $a_{i,j}$ of a matrix $A$ is said to be "matching" if $a_{i,j}$ is non-zero and $a_{j,i}$ is also non-zero.
\end{itemize}

\subsection{Value Based Features}
\begin{itemize}
    \item \emph{statistical features of matrix elements:} The minimum, maximum, mean, and standard deviation of all the matrix elements.
    \item \emph{statistical features of row and column averages:} The minimum, maximum, mean, and standard deviation of the average element for each row, and the same for the column averages.
    \item \emph{statistical features of row and column standard deviations:} The minimum, maximum, mean, and standard deviation of the standard deviation of elements in each row, and the same for the column standard deviations.
    \item \emph{all statistical features for non-zero elements:} Similar to the features described in the above items "statistical features of matrix elements," "statistical features of row and column averages," and "statistical features of row and column standard deviations," but incorporating only the relevant non-zero elements into the computation of each feature.
    \item \emph{statistical features of row and column sums:} The minimum, maximum, mean, and standard deviation of the sum of elements in each row, and the same for the column sums.
    \item \emph{statistical features of diagonal elements:} The mean and standard deviation of the elements in the main matrix diagonal.
    \item \emph{statistical features of upper triangular and lower triangular elements:} The mean and standard deviation of the elements in the upper triangular part of the matrix, and the same for the lower triangular elements.
    \item \emph{norm features:} The one-norm, two-norm, infinity-norm, and Frobenius norm of the matrix.
    \item \emph{symmetric Frobenius norm:} The Frobenius norm of the symmetric part of the matrix, where the symmetric part of a matrix $A$ is defined as $1/2*(A + A^T)$.
    \item \emph{asymmetric Frobenius norm:} The Frobenius norm of the asymmetric part of the matrix, where the asymmetric part of a matrix $A$ is defined as $1/2*(A - A^T)$.
\end{itemize}

\subsection{Diagonal Based Features}
\begin{itemize}
    \item \emph{lower and upper bandwidth:} Measured by finding the last diagonal below or above the main diagonal that contains non-zero entries. Specifically, the lower bandwidth is the smallest number $B_{low}$ such that $a_{i,j} = 0$ whenever $i - j > B_{low}$ for a matrix $A$. The upper bandwidth is the smallest number $B_{up}$ such that $a_{i,j} = 0$ whenever $j - i > B_{up}$.
    \item \emph{statistical features of column width (bandwidth):} Average and maximum column width of the matrix, where the width of a column is defined as the difference between the row indices of the lowest-row non-zero element and highest-row non-zero element in that column.
    \item \emph{distance from diagonal:} The average and standard deviation of the distance from each non-zero element to the diagonal. This is defined as $|i-j|$ for a non-zero element $a_{i,j}$ of a matrix $A$.
    \item \emph{difference from diagonal:} The average and standard deviation of the difference between each non-zero element and its corresponding diagonal element in the same row.
    \item \emph{row maximum difference from diagonal:} The average and standard deviation of the differences between each maximum row element and its respective diagonal value in the same row.
    \item \emph{diagonal dominance degree:} The percentage of diagonally dominant columns and the percentage of diagonally dominant rows. A column is said to be diagonally dominant if the absolute value of the diagonal element in that column is greater than the sum of absolute values of the non-diagonal elements in that column, and similarly for rows.
    \item \emph{diagonal value rate:} The ratio of the minimum to maximum non-zero elements in the main diagonal of the matrix.
\end{itemize}

\subsection{Condition Number Features}
\begin{itemize}
    \item \emph{Gershgorin disks estimation:} The Gershgorin Disk Theorem \cite{garren_bounds_1968} is used to determine a real upper and lower bound, $\lambda_{max}$ and $\lambda_{min}$, respectively, on the magnitude of the matrix eigenvalues. Both are used as features, along with their ratio $\lambda_{max}/\lambda_{min}$, which acts as an estimate of the condition number. An additional feature, the overlap, is calculated as the total overlapping distance between every two Gershgorin disks on the real axis.
    \item \emph{Cassini ovals estimation:} The Cassini Ovals Theorem \cite{garren_bounds_1968}, an alternative to Gershgorin Disks, is used to determine $\lambda_{max}$, $\lambda_{min}$, and their ratio.
    \item \emph{condition number:} The actual condition number, $\kappa$, of the matrix. This feature is only used in certain models for comparison with other feature sets and is not for use in a general solution due to the computational complexity of calculating the condition number.
\end{itemize}

\end{document}